\documentclass[11pt]{article}
\usepackage{graphics,epsfig}
\usepackage[]{graphicx}
\usepackage{amssymb}
\usepackage{latexsym}

\begin{document}
\title{Q-spaces and the foundations of quantum mechanics}
\author{{\sc Graciela
Domenech$^{1}$}\thanks{%
Fellow of the Consejo Nacional de Investigaciones Cient\'{\i}ficas y
T\'ecnicas (CONICET) - Argentina}, \ {\sc Federico Holik$^{1}$} \ and \ {\sc D\'ecio
Krause$^{2}$}\thanks{Fellow of the Conselho Nacional de Desenvolvimento Cient\'{\i}fico e Tecnol\'ogico (CNPq) - Brazil}}

\maketitle

\begin{center}
\begin{small}
\noindent ${}^{1}$ Instituto de Astronom\'{\i}a y F\'{\i}sica del Espacio (IAFE)\\
Casilla de Correo 67, Sucursal 28, 1428 Buenos Aires, Argentina\\ and \\
\noindent ${}^{2}$ {Departamento de Filosofia - Universidade Federal de Santa Catarina} \\
{P.O.Box 476, 88040-900 Forian\'opolis, SC - Brazil}
\end{small}
\end{center}


\begin{abstract}
\noindent Our aim in this paper is to take quite seriously Heinz
Post's claim that the non-individuality and the indiscernibility of
quantum objects should be introduced \emph{right at the start}, and
not made a posteriori by introducing symmetry conditions. Using a
different mathematical framework, namely, quasi-set theory, we avoid
working within a label-tensor-product-vector-space-formalism, to use
Redhead and Teller's words, and get a more intuitive way of dealing
with the formalism of quantum mechanics, although the underlying
logic should be modified. Thus, this paper can be regarded as a
tentative to follow and enlarge Heinsenberg's suggestion that new
phenomena require the formation of a new ``closed" (that is,
axiomatic) theory, coping also with the physical theory's underlying
logic and mathematics.
\end{abstract}

\begin{small}
\centerline{\em Key words:  quasi-sets, particle number, Fock space,
quantum indistinguishability.}
\end{small}

\newtheorem{theo}{Theorem}[section]
\newtheorem{definition}[theo]{Definition}
\newtheorem{lem}[theo]{Lemma}
\newtheorem{prop}[theo]{Proposition}
\newtheorem{coro}[theo]{Corollary}
\newtheorem{exam}[theo]{Example}
\newtheorem{rema}[theo]{Remark}{\hspace*{4mm}}
\newtheorem{example}[theo]{Example}
\newtheorem{axiom}[theo]{Axiom}
\newcommand{\proof}{\noindent {\em Proof:\/}{\hspace*{4mm}}}
\newcommand{\qed}{\hfill$\Box$}
\newcommand{\ninv}{\mathord{\sim}} 
\newcommand{\Q}{$\mathfrak{Q}$}
\newcommand{\igual}{\stackrel{\tiny{\mathrm{def}}}{=}}

\newcommand{\lra}{\leftrightarrow}

\bibliography{pom}

\begin{thebibliography}{10}

\bibitem{bal00} Ballentine, L.\ E.: 2000, \emph{Quantum Mechanics: A Modern Development}, Singapore, World Scientific.

\bibitem{bec08} Becker, J.: 2008, \emph{Topics on Quasi-Set Theory and on its Applications to the Philosophy of Quantum Physics} (in Portuguese), MSc dissertation, Federal University of Santa Catarina.

\bibitem{bok06} Bokulich, A.: 2006, ``Heisenberg Meets Kuhn: Closed
Theories and Paradigms'', \emph{Philosophy of Science}, 73, 90–107.

\bibitem{coskra94} da Costa, N.\ C.\ A. and  Krause, D.: 1994, ``Schr\"{o}dinger logics'',
\textit{Studia Logica} 53 (4),  533-550.

\bibitem{coskra97} da Costa, N.\ C.\ A. and  Krause, D.: 1997, ``An intensional Schr\"{o}dinger logic'',
\textit{Notre Dame Journal of Formal Logic} 38 (2),
 179-194.

\bibitem{coskra07} da Costa, N.\ C.\ A.\ and Krause, D.: 2007, ``Logical and Philosophical Remarks
on Quasi-Set Theory'', \emph{Logic Journal of the Interest Group in Pure
and Applied Logics}, 15, 1-20.

\bibitem{daltor95} Dalla Chiara, M.\ L.\ and Toraldo di Francia, G.: 1995, ``Identity Questions from
 Quantum Theory'', in Gavroglu, K. \emph{et.\ al.}, (eds.), \emph{Physics,
 Philosophy and the Scientific Community}, Dordrecht, Kluwer Academic
 Publishers, 39-46.

\bibitem{dalgiukra98}  Dalla Chiara, M.\ L., Giuntini, R.\ and Krause, D.: 1998,
``Quasiset Theories for Microobjects: A Comparision'', in
Castellani, E. (ed.), {\it Interpreting bodies: Classical and
quantum objects in modern physics}, Princeton
 University Press, Princeton.

\bibitem{domhol07} Domenech, G.\ and Holik, F.: 2007, ``A discussion on particle number and quantum indistinguishability", \emph{Found. Phys.} 37 (6), 855-878.

\bibitem{fal07} Falkenburg, B.: 2007, \textit{Particle Metaphysics: A Critical Account of Subatomic Reality}, Springer

\bibitem{frekra03} French, S.\ and Krause, D.: 2003, ``Quantum Vagueness", \textit{Erkenntnis} 59,  97-124.

\bibitem{frekra06} French, S.\ and Krause, D.: 2006, {\it Identity  in Physics: A Historical, Philosophical,
and Formal Analysis}, Oxford, Oxford University Press.

\bibitem{hal63} Halmos P.: 1963, {\it Naive Set Theory}, D. Van Nostrand
Company.

\bibitem{halcli01}  Halvorson, H. P. and  Clifton, R.: 2001,  ''Are Rindler quanta real? Inequivalent particle concepts
in quantum field theory'', \emph{British Journal for the Philosophy of Science}, 52, 417-470.

\bibitem{hei79} Heisenberg, W.: 1979,  ``Recent Changes in the Foundations of Exact Science", in F. C. Hayes (trans.), \emph{Philosophical Problems of Quantum Physics}, Woodbridge, CT: Ox Bow Press,
11–26.

\bibitem{kra92} Krause, D.: 1992, ``On a quasi set theory", \textit{Notre Dame
J.\ of Formal Logic} 33, 402-411.


\bibitem{kra03} Krause, D.: 2003, ``Why quasi-sets?'', \emph{Boletim da Sociedade Paranaense de
Matem\'atica} {\bf 20}, 73-92.

\bibitem{krasanvol99} Krause, D.,  Sant'Anna, A.\ S.\  and  Volkov, A.\ G.: 1999, ``Quasi-set
theory for bosons and fermions: quantum distributions",
\textit{Foundations of Physics Letters} \textbf{12} (1), pp.\
51-66.

\bibitem{krasansar05} Krause, D., Sant'Anna, A.\ S.\ and Sartorelli, A.: 2005, ''A critical study on the
 concept of identity in Zermelo-Fraenkel like axioms and its relationship with quantum statistics'',
 \emph{Logique $\&$ Analyse}, 189-192, 231-260.

\bibitem{lei95} Leibinz, G.\ W.: 1995, \emph{Philosophical Writings}, Ed.\ by G.\ H.\ R.\ Parkinson, London, Everyman.

\bibitem{man76} Manin, Yu. I.: 1976, ``Mathematical Problems I: Foundations'', in Browder, F.\ E.\ (ed.): 1976, \textit{Mathematical Problems Arising
from Hilbert Problems}, Proceedings of Symposia in Pure
Mathematics, Vol.\ XXVIII, Providence, American Mathematical
Society, 36.

\bibitem{man77} Manin, Yu. I.: 1977, \emph{A course in mathematical logic}, Springer-Verlag. 

\bibitem{mit98} Mittelstaedt, P.: 1998, {\it The interpretation of quantum mechanics and the
measurement process}, Cambridge University Press.

\bibitem{pos63} Post, H.: 1963, ``Individuality and physics", \textit{The Listener}, 10 October,
534-537, reprinted in \textit{Vedanta for East and West} \textbf{132}, 1973, 14-22.

\bibitem{redtel91} Redhead, M.\ and Teller, P.: 1991, ``Particles, particle labels,
and quanta: the toll of unacknowledged metaphysics", {\it Foundations of Physics}
{\bf  21}, pp.\ 43-62.

\bibitem{redtel92} Redhead, M.\ and Teller, P.: 1992, ``Particle labels and the
theory of indistinguishable particles in quantum mechanics", {\it British Journal
for  the Philosophy of Science\/}  {\bf 43}, pp.\ 201-218.
pp.\ 14-22.

\bibitem{rob73} Robertson, B.: 1973, ``Introduction to field operators in quantum mechanics", \emph{Amer. J. Phys.} 41 (5), 678-690

\bibitem{pen89} Penrose, R.: 1989 {\it The emperor´s new mind}, Oxford, Oxford Un.\ Press.

\bibitem{san05} Sant'Anna, A.: 2005, ``Labels for non individuals?'', \emph{Found. Phys.
Lett.} 18, 0894-9875.

\bibitem{sch52} Schr\"odinger, E.: 1952,  \textit{Science and Humanism},
Cambridge Un.\ Press, Cambridge.

\bibitem{sch98} Schr\"{o}dinger, E.: 1998, ``What is an elementary particle?'', reprinted in Castellani, E.
(ed.), {\it Interpreting bodies: classical and quantum objects in
modern physics}, Princeton Un. Press, 197-210.

\bibitem{tel95} Teller, P.: 1995, \textit{An Interpretative Introduction to Quantum Field
Theory}, Princeton, Princeton University Press.

\bibitem{tor81} Toraldo di Francia,  G.: 1981, \textit{The Investigation of the Physical World}, Cambridge Un.\
Press.

\bibitem{wey49} Weyl, H.: 1949, {\it Philosophy of mathematics and natural science}, Princeton,
 Princeton University Press.
\end{thebibliography}


\section{Introduction}
\hfill{
\parbox{4.0in}
{ \hrulefill

 {\footnotesize
``The transition in science from previously investigated fields of experience
to new ones will never consist simply of the application of
already known laws to these new fields. On the contrary, a really new
field of experience will always lead to the crystallization of a new
system of scientific concepts and laws ($\ldots$). The advance from the
parts already completed to those newly discovered, or to be newly
erected, demands each time an intellectual jump, which cannot be
achieved through the simple development of already existing knowledge."
\, W.\ Heisenberg \cite[p.\ 25]{hei79}, quoted from \cite{bok06}

\hrulefill} }}
\vspace{3mm}

In his paper ``Individuality in physics" (\cite{pos63}), Heinz
Post claimed that ``[p]articles are non-individual in modern theory
($\ldots$) [and] non-individuality has to be introduced right at the
start". Our aim in this paper is to take quite seriously Heinz
Post's claim. Usually, the way of dealing with indiscernible objects
within the scope of classical logic and mathematics is by
restricting them to certain structures, such that the relations and
functions of the structure are not sufficient to individuate them.
Saying in other words, such structures are not rigid, in the sense
that there are automorphisms other than the identity function. For
instance, within the additive group of the integers $\mathcal{Z} =
\langle \mathbb{Z}, + \rangle$, there is no way of distinguishing
between two integers $n$ and $-n$, for the function $f(x) = -x$ is
an automorphism of the structure. But in standard mathematics, such
as in that one that can be built in Zermelo-Fraenkel set theory with
the axiom of foundation (ZF), which we can assume bases all physical
theories and which we shall identify with ``classical mathematics",
any structure can be extended to a \emph{rigid} structure, that is,
to a structure whose only automorphism is the identity function.
That means that, \emph{outside} the group $\mathcal{Z}$, for
instance in the rigid extended structure $\mathcal{Z}' = \langle
\mathbb{Z}, +, < \rangle$, we of course can distinguish, say,
between $3$ and $-3$ (for $-3 < 3$  but not reciprocally). The fact
that any structure (built in ZF) can be extended to a rigid
structure makes the indiscernibility of the objects something quite
artificial. That means that, although we can deal with certain
objects \emph{as if they were} indiscernible, from ``outside" of
these structures these objects are not indiscernible, for they can
be individualized in the extended rigid structures. In particular,
in the ``whole ZF", that is, in the well-founded ``structure"
$\mathcal{V} = \langle V, \in \rangle$, where $V$ is the von Neumann
well founded universe and $\in$ is the membership relation, every
representable object is an \emph{individual}, in the sense that it
obeys the laws of identity of classical (first or higher order)
logic. In other words, any object $a$ can be distinguished from  any
other object $b$, say by the fact that it (and it only) belongs to
its singleton $\{a\}$ (which can be identified --in extensional
contexts-- with the property ``being identical to $a$", namely,
$P(x) \igual x=a$).

In quantum physics, which is of course standardly build within
classical mathematics (thus encompassing classical logic as well),
the strategy is quite similar to that one mentioned above. When a
system composed of particles of the same kind is considered, we
start by labeling the particles we are dealing with,\footnote{The
term ``particle" would not be taken literally as denoting neither
tinny entities like usual bodies, nor entities as described by
classical physics (for the differences, see \cite[chap.\ 6]{fal07}).
For our argumentation, it does not matter what sort of entities they
``really" are, but only that, yet far from the na\"{\i}ve idea
associated with this word, ``the particle concept [of particle] was
not given up. After the quantum revolution, physicists still speak
of particles" \cite[p.\ 209]{fal07}. We shall take this quotation ever
in mind in this paper. An analogous consideration holds for the term
``system".} say by naming them ``1", ``2" and so on, and then
consider the relevant Hilbert spaces $\mathcal{H}_1$,
$\mathcal{H}_2$, etc., for each particle and, for the join system,
we take the tensorial product $\mathcal{H} = \bigotimes_{i \in I}
\mathcal{H}_i$. The base vectors $| \alpha_1 \rangle \otimes
|\alpha_2\rangle \otimes \ldots$, or simply $|
\alpha_1\alpha_2\ldots\rangle$ must be so that for any permutation
operator $P_{ij}$, which intuitively speaking exchanges the labels
(hence, the particles) $i$ and $j$, we have
$$P_{ij}| \alpha_1\ldots\alpha_i\ldots\alpha_j\ldots\rangle = \pm| \alpha_1\ldots\alpha_j\ldots\alpha_i\ldots\rangle,$$

\noindent that is, the state of indistinguishable particles is left
\emph{invariant by permutations}. The plus sign stands for bosons,
and the minus sign for fermions. This
\emph{labeled-tensor-product-Hilbert-space formalism} (LTPHSF), as
called by Redhead and Teller \cite{redtel91}, \cite{redtel92},
\cite{tel95}, requires that symmetry conditions of this kind are
introduced: we start from \emph{individuals}, say by choosing a
vector basis $\{|\alpha_i\rangle\}$ for a suitable Hilbert space,
which serves as a kind of label to the particle. There is no scape.
In order to talk of objects, we need to refer to them in some way
(Toraldo di Francia says that our languages --including those of
science-- are ``objectual" \cite{tor81}), and this perhaps is due to
the deficiencies of the language employed, taken from classical
physics, which, as said Schr\"odinger, ``gets off on the wrong
foot'' by initially assigning particle labels and then permuting
them before extracting combinations of appropriate symmetry
\cite{kra03}.  It is a challenge to find a suitable language that
enables us to speak of indiscernible objects without making such
first hypothesis about their individuality. Really, once
indiscernibility lies in the core of quantum assumptions, perhaps we
need to agree with Yuri Manin when he says that ``quantum mechanics
has not its \emph{own} language" (\cite[p.\ 84]{man77}), for
(ideally) a suitable language would refer to quantum objects without
identifying them as individuals (this of course poses difficulties
also for the use of quantifiers in physics). The introduction of
symmetry conditions, say by choosing symmetric and anti-symmetric
functions (or vectors), are necessary devices in the formalism. In
the above mentioned papers, Redhead and Teller point to various
puzzles caused by such an assumption, and propose the use of the
Fock space formalism instead. We shall turn to their claim later.
Yet, ``for all practical purposes" (in Bell's words), LTPSF works
quite well, as present day physics exemplifies, but rigorously
speaking, we see that there is an enormous gap, for we are
considering individuals at the start, and them ``make" them
non-individuals by some mathematical trick, like by ``forgetting"
that they are individuals and considering only certain quantities of
them in each situation (this strategy was called ``the Weyl
strategy" in \cite{frekra06}, where there are further philosophical
discussions). Thus, to pursue Post's claim seems to be relevant.

We guess that quasi-set theory provides a language for dealing with
collections of indiscernible elements right from the start.  In this
theory, the notion of indistinguishability (or indiscernibility) is
taken as a primitive notion, and a definite concept of identity is
restricted so that there may exist objects that are indiscernible
without turning to be identical (to be \emph{the same} object).
Thus, the theory is non-Leibnizian, for his principle of the
identity of indiscernibles is not valid in general: entities can
have all the same properties without turning to be identical. We
sketch the quasi-set theory \Q\ in the next section (for further
details, see \cite[Chap.\ 7]{frekra06}). In \ref{s:Quaiset} we
review general notions on quasi-set theory. In \ref{s:Q-space}, we
use the non classical part of quasi-set theory to construct a vector
spaces with inner product, which is adequate to deal with bosons and
fermions. In section \ref{s:fock}, we use this space to formulate
quantum mechanics of indistinguishable particles without appealing
to intermediate indexations. Finally, we expose our conclusions in
\ref{s:discussion}.

\section{Quasi-set theory}\label{s:Quaiset}
We recall here some notions of quasi-set theory that will play an
important role in what follows (for further details, see
\cite[Chap.\ 7]{frekra06}). We shall not present all the postulates and
definitions of the theory, but just revise the main ideas and
results which interest us here. Intuitively speaking a quasi-set is
a collection of indistinguishable (but not identical) objects. This
of course is not a strict ``definition" of a quasi-set, acting more
or less as Cantor's ``definition" of a set as ``any collection into
a whole $M$ of definite and separate [that is, distinguishable]
objects $m$ of our intuition or our thought" (see the discussion in
\cite[\S 6.4]{frekra06}), giving no more than an intuitive account of the
concept.

The quasi-set theory  \Q\ was conceived to handle collections of
indistinguishable objects, and was motivated by some considerations
taken from quantum physics, mainly in what respects Schr\"odinger's
idea that the concept of identity would not be applied to elementary
particles \cite[pp.\ 17-18]{sch52}. Of course the theory can be
developed independently of any formulation of quantum mechanics, but
here we shall have this motivation always in mind. Our way to deal
with indistinguishability is by assuming that expressions like $x=y$
are not always well formed. We express that by saying that the
concept of identity \emph{does not apply} to the entities denoted by
$x$ and $y$ when they ``refer" to quantum objects. Due to the lack
of sense in applying the concept of identity to certain elements,
informally, a quasi-set (qset), that is, a collection involving such
objects, may be such that its elements cannot be identified by
names, counted, ordered, although there is a sense in saying that
these collections have a cardinal (not defined by means of ordinals,
as usual --but see below). But we aim at to keep standard
mathematics intact, so the theory is developed in a way that ZFU
(and hence ZF, perhaps with the axiom of choice, ZFC) is a subtheory
of \Q. In other words, the theory is constructed so that it extends
standard Zermelo-Fraenkel with \textit{Urelemente} (ZFU) set theory;
thus standard sets (of ZFU) can be viewed as particular qsets (that
is, there are qsets that have all the properties of the sets of ZFU;
the objects in \Q\ corresponding to the \textit{Urelemente} of ZFU
are termed $M$-atoms). These objects will be called \Q-sets, or just
\emph{sets} when there will be no confusion.  But quasi-set theory
encompasses another kind of \textit{Urelemente}, the $m$-atoms, to
which the standard theory of identity does not apply (that is,
expressions like $x = y$ are not well formed if either $x$ or $y$
denote $m$-atoms). Thus, we can say that \Q-sets are qsets whose
transitive closure (defined as usual --see below) does not contain
$m$-atoms (in other words, they are ``constructed" in the
"classical" part of the theory --see Fig.\ 1).


\begin{figure}[h]
\setlength{\unitlength}{1mm} \centering
\begin{picture}(80,70)

\put(15,5){\line(-1,3){15}}

\put(15,5){\line(1,0){50}}

\put(65,5){\line(1,3){15}}

\put(65,5){\line(-1,3){15}}

\put(37,5){\line(-1,3){15}}

\put(80,46){\Large{$\mathsf{Q}$}}

\put(17,6){\scriptsize{$m$-atoms}}

\put(40,6){\scriptsize{$M$-atoms}}

\put(35,28){\scriptsize{The ``classical part" of \Q}}

\put(8,40){\scriptsize{\textit{pure} qsets}}

\put(67,5){$\emptyset$}

\multiput(65,5)(0,1){46}{\line(0,1){.3}}

\put(66,46){$On$}

\put(55,40){\scriptsize{copies of \textsf{ZF}-sets}}

\put(28,40){\scriptsize{copies of \textsf{ZFU}-sets}}

\end{picture}

 \caption{\small {The quasi-set universe: $\mathsf{Q}$ is a ``model" of \Q.}}\label{qsetuniverse}
\end{figure}

When \Q\ is used in connection with quantum physics, these $m$-atoms
are thought of as representing quantum objects (henceforth,
\emph{q-objects}), and not necessarily they are `particles'; waves
or perhaps even strings (and whatever `objects' sharing the property
of indistinguishability of pointlike elementary particles) can be
also be values of the variables of \Q\ (see \cite[Chap.\ 6]{fal07}
for an account on the various ways to understand the word
``particle" in connection to quantum physics). The lack of the
concept of identity for the $m$-atoms makes them
\emph{non-individuals} in a sense, and it is mainly (but not only)
to deal with collections of $m$-atoms that the theory was conceived.
So, \Q\ is a theory of generalized collections of objects, involving
non-individuals. For details about \Q\ and about its historical
motivations, see \cite[Chap.\ 7]{frekra06}.

In order to distinguish between \Q-sets and qsets that have
$m$-atoms in their transitive closure, we write (in the
metalanguage) $\{x : \varphi(x)\}$ for the former and $[x :
\varphi(x)]$ for the latter. In \Q, the so called `pure' qsets have
only q-objects as elements (although these elements may be not
always indistinguishable from one another), and to them it is
assumed that the usual notion of identity cannot be applied (that
is, let us recall, $x=y$, so as its negation, $x \not= y$, are not a
well formed formulas if either $x$ or $y$ stand for q-objects).
Notwithstanding, there is a primitive relation $\equiv$ of
indistinguishability having the properties of an equivalence
relation, and a concept of \textit{extensional identity}, not
holding among $m$-atoms, is defined and has the properties of
standard identity of classical set theories. More precisely, we
write $x =_E y$ ($x$ and $y$ are extensionally identical) iff they
are both qsets having the same elements (that is, $\forall z (z \in
x \lra z \in y)$) or they are both $M$-atoms and belong to the same
qsets (that is, $\forall z (x \in z \lra y \in z)$). From now on, we
shall use the symbol ``='' for the extensional equality, except when
explicitly mentioned.

Since the elements of a qset may have properties (and satisfy
certain formulas), they can be regarded as
\textit{indistinguishable} without turning to be \textit{identical}
(that is, being \textit{the same} object), that is, $x \equiv y$
does not entail $x=y$. Since the relation of equality (and the
concept of identity) does not apply to $m$-atoms, they can also be
thought of as entities devoid of individuality. We remark further
that if  the `property' $x=x$ (to be identical to itself, or
\textit{self-identity}, which can be defined for an object $a$ as
$I_a(x) \igual x = a$) is included as one of the properties of the
considered objects, then the so called Principle of the Identity of
Indiscernibles (PII) in the form $\forall F (F(x) \leftrightarrow
F(y)) \to x=y$ is a theorem of classical second order logic, and
hence there cannot be indiscernible but not identical entities (in
particular, non-individuals). Thus, if self-identity is  linked to
the concept of non-individual, and if quantum objects are to be
considered as such, these entities fail to be self-identical, and a
logical framework to accommodate them is in order (see \cite{frekra06}
for further argumentation).

We have already discussed at length in the references given above
(so as in other works) the motivations to build a quasi-set theory,
and we shall not return to these points here,\footnote{But see
\cite{kra92}, \cite{coskra94}, \cite{coskra97}, \cite{kra03}, \cite{coskra07},  \cite{frekra03}, \cite{kra96},
\cite{krasansar05}, \cite{frekra06}.} but before to continue we would
like to make some few remarks on a common misunderstanding about PII
and quantum physics. People generally think that spatio-temporal
location is a sufficient condition for individuality. Thus, two
electrons in different locations \emph{are} discernible, hence
\emph{distinct individuals}. Leibniz himself prevented us about this
claim (yet not directly about quantum objects of course), by saying
that ``it is not possible for two things to differ from one another
in respect to place and time alone, but that is always necessary
that there shall be some other internal difference" \cite{lei95}.
Leaving aside a possible interpretation for the word `internal', we
recall that even in quantum physics, fermions obey the Pauli
Exclusion Principle, which says that two fermions (yes, they `count'
as more than one) cannot have all their quantum numbers (or
`properties') in common. Two electrons (which are fermions), one in
the South Pole and another one in the North Pole, \emph{are not
individuals in the standard sense} (and we can do that without
discussing the concepts of space and time). Here, by an individual
we understand an object that obeys the classical theory of identity
of classical (first or higher order) logic (extensional set theory
included). In fact, we can say that the electron in the South Pole
is described by the wave function $\psi_S(x)$, while the another one
is described by $\psi_N(x)$ (words like `another' in the preceding
phrase are just ways of speech, done in the informal metalanguage).
But the wave function of the joint system is given by
$\psi_{SN}(x_{1},x_{2})=\psi_S(x_{1})\psi_N(x_{2})-
\psi_N(x_{1})\psi_S(x_{2})$ (the function must be anti-symmetric in
the case of fermions, that is, $\psi_{NS}(x_{1},x_{2})= -
\psi_{NS}(x_{2},x_{1})$), a superposition of the product wave
functions $\psi_S(x_{1})\psi_N(x_{2})$ and
$\psi_S(x_{2})\psi_N(x_{1})$. Such a superposition cannot be
factorized. Furthermore, in the quantum formalism, the important
thing is the square of the wave function, which gives the joint
probability density; in the present case, we have
$||\psi_{SN}(x_{1},x_{2})||^2 = ||\psi_S(x_{1})\psi_N(x_{2})||^2 +
||\psi_S(x_{2})\psi_N(x_{1})||^2 -
2\mathrm{Re}(\psi_S(x_{1})\psi_N(x_{2})\psi_S(x_{2})^{\ast}\psi_N(x_{1})^{\ast})$.
This last `interference term' (though vanishing at large distances),
cannot be dispensed with, and says that nothing, not even \emph{in
mente Dei}, can tell us which is the particular electron in the
South Pole (and the same happens for the North Pole). As far as
quantum physics is concerned, they really and truly have no identity
in the standard sense (and hence they have not \emph{identity} at
all).

\subsection{The basic ideas of quasi-set theory}
Quasi-sets are the collections obtained by applying ZFU-like
(Zermelo-Fraenkel plus \textit{Urelemente}) axioms to a basic domain
composed of $m$-atoms, $M$-atoms and aggregates of them. The theory
still admits a primitive concept of quasi-cardinal which intuitively
stands for the `quantity' of objects in a collection. This is made
so that certain quasi-sets $x$ (in particular, those whose elements
are q-objects) may have a quasi-cardinal, written $qc(x)$, but not
an ordinal.  It is also possible to define a translation from the
language of ZFU into the language of \Q\ in such a way so that there
is a `copy' of ZFU in \Q\ (the `classical' part of $\mathfrak Q$).
In this copy, all the usual mathematical concepts can be defined
(inclusive the concept of ordinal for the \Q-sets), and the \Q-sets
turn out to be those quasi-sets whose transitive closure (this
concept is like the usual one) does not contain
$m$-atoms.\footnote{So, we can make sense to the primitive concept
of quasi-cardinal of a quasi-set $x$  as being a cardinal defined in
the `classical' part of the theory. The reason to take the concept
of quasi-cardinal as a primitive concept will appear below, when we
make reference to the distinction between cardinals and ordinals.
The first two authors of this paper have defined the quasi-cardinal
for finite qsets; see \cite{domhol07}. Independently, Becker has
extending this idea with other considerations \cite{bec08}.}

To understand the basic involved ideas, let us consider the three
protons and the four neutrons in the nucleus of a $^\textsf{7}{\rm
Li}$ atom. As far as quantum mechanics goes, nothing distinguishes
these \textit{three} protons. If we regard these protons as forming
a quasi-set, its quasi-cardinal should be 3, and there is no
apparent contradiction in saying that there are also 3 subquasi-sets
with 2 elements each, despite we can't distinguish their elements,
and so on. So, it is reasonable to postulate that the quasi-cardinal
of the power quasi-set of $x$ is $2^{qc(x)}$. Whether we can
distinguish among these subquasi-sets is a matter which does not
concern logic.

In other words, we may consistently (with the axiomatics of \Q)
reason as if there are three entities in our quasi-set $x$, but $x$
must be regarded as a collection for which it is not possible to
discern its elements as individuals. The theory does not enable us
to form the corresponding singletons. The grounds for such kind of
reasoning has been delineated by Dalla Chiara and Toraldo di Francia
as partly theoretical and partly experimental. Speaking of electrons
instead of protons, they note that in the case of the helium atom we
can say that there are two electrons because,
\textit{theoretically}, the appropriate wave function depends on six
coordinates and thus ``we can therefore say that the wave function
has the same degrees of freedom as a system of two classical
particles".\footnote{Op.\ cit., p.\ 268. This might be associated to
the legacy of  Schr\"odinger, who says that this kind of formulation
``gets off on the wrong foot'' by initially assigning particle
labels and then permuting them before extracting combinations of
appropriate symmetry \cite{sch98}.} Dalla Chiara and Toraldo di
Francia have also noted that, ``[e]xperimentally, we can ionize the
atom (by bombardment or other means) and extract two separate
electrons $\ldots$'' (ibid.).

Of course, the electrons can be counted as two only at the moment of measurement; as
soon as they interact with other electrons (in the measurement apparatus, for
example) they enter into entangled states once more. It is on this basis that one
can assert that there are two electrons in the helium atom or six in the 2p level of
the sodium atom or (by similar considerations) three protons in the nucleus of a
$^\textsf{7}{\rm Li}$ atom (and it may be contended that the `theoretical' ground
for reasoning in this way also depends on these experimental considerations,
together with the legacy of classical metaphysics). On this basis it is stated the
axiom of `weak extensionality' of \Q, which says that those quasi-sets
that have the same quantity of elements of the same sort (in the sense that they
belong to the same equivalence class of indistinguishable objects) are
indistinguishable.

This axiom has interesting consequences. As we have said, there is
no space here for the details, but let us  mention just one of them
which is related to the above discussion on the non observability of
permutations in quantum physics, which is one of the most basic
facts regarding indistinguishable quanta. In standard set theories,
if $w \in x$, then of course $(x - \{w \}) \cup \{z\} = x$ iff $z =
w$. That is, we can 'exchange' (without modifying the original
arrangement) two elements iff they are \textit{the same} elements,
by force of the axiom of extensionality. But in \Q\ we can prove the
following theorem, where $z'$ (and similarly $w'$) stand for a
quasi-set with quasi-cardinal 1 whose only element is
indistinguishable from $z$ (respectively, from $w$ --the reader
shouldn't think that this element \textit{is identical to either}
$z$ or $w$, for the relation of equality doesn't apply here; the set
theoretical operations can be understood according to their usual
definitions):

{\sf [Unobservability of Permutations]}\label{unobservabilty}
 Let $x$ be a finite quasi-set such that
$x$ does not contain all indistinguishable from $z$, where $z$ is an $m$-atom such
that $z \in x$. If $w \equiv z$ and $w \notin x$, then there exists $w'$ such that
 $$(x - z') \cup w' \equiv x$$

Supposing that $x$ has $n$ elements, then if we `exchange' their
elements $z$ by correspondent indistinguishable elements $w$ (set
theoretically, this means performing the operation $(x - z') \cup
w'$), then the resulting quasi-set remains
\textit{indistinguishable} from the original one. In a certain
sense, it is not important whether we are dealing with $x$ or with
$(x - z') \cup w'$. This of course gives a 'set-theoretical' sense
to the following claim made by Roger Penrose:

\begin{quotation}
``[a]ccording to quantum mechanics, any two electrons must
necessarily be completely identical [in the physicist's jargon, that
is, indistinguishable], and the same holds for any two protons and
for any two particles whatever, of any particular kind. This is not
merely to say that there is no way of telling the particles apart;
the statement is considerably stronger than that. If an electron in
a person's brain were to be exchanged with an electron in a brick,
then the state of the system would be \textit{exactly the same
state} as it was before, not merely indistinguishable from it! The
same holds for protons and for any other kind of particle, and for
the whole atoms, molecules, etc. If the entire material content of a
person were to be exchanged with the corresponding particles in the
bricks of his house then, in a strong sense, nothing would be
happened whatsoever. What distinguishes the person from his house is
the \textit{pattern} of how his constituents are  arranged, not the
individuality of the constituents themselves'' \cite[p.\
32]{pen89}.
\end{quotation}

Within {$\mathfrak Q$} we can express that `permutations are not
observable', without necessarily introducing symmetry postulates,
and in particular to derive `in a natural way' the quantum
statistics (see \cite{krasanvol99}, \cite[Chap.\ 7]{frekra06}).

\section{The \Q-space}\label{s:Q-space}
In the standard formulation of quantum mechanics, pure states of
quantum systems are represented by normalized to unit vectors in a
Hilbert space. In the case of identical particles, the vectors
representing their states are symmetrized or antisymmetrized, as
mentioned above. In this section, we will use \Q\ to construct a
vector space, which we will call \Q-\emph{space}, in which the
states are defined without labeling particles for they are
represented by $m$-atoms. The structure of this space will result
analogous to that of the Fock-space.

\subsection{Motivation}\label{motivation}
Let us analyze with a deeper detail how quantum mechanics deals with
a system of two indistinguishable particles, just to introduce some
notation and to motivate our construction. Recall that the usual
construction of a vector space --and of the whole formalism of
quantum mechanics-- makes use of set theory, which presupposes the
individuality and distinguishability of the elements of any set.
First the Hilbert space $\mathcal{H} =
\mathcal{H}_1\bigotimes\mathcal{H}_2$ is constructed up from the one
particle spaces $\mathcal{H}_1$ and $\mathcal{H}_2$. We use Dirac
notation for simplicity. Let $\{|\alpha\rangle\}$ be a basis set of
$\mathcal{H}_i$. Then, $\{|\alpha \rangle \otimes |\beta\rangle\}$
is a basis for $\mathcal{H}$. $\alpha$ and $\beta$ run over all
possible values of the corresponding physical magnitudes and it is
understood that the first ket corresponds to the particle labeled
``1" and the second to the one labeled ``2".

The scalar product of any two basis vectors is given by:
\begin{equation}
(\langle\alpha|\otimes\langle\beta|)(|\alpha'\rangle\otimes|\beta'\rangle)=
\langle\alpha|\alpha'\rangle\langle\beta|\beta'\rangle
\end{equation}
\noindent and, in general, the scalar product between two product
vectors $|\psi\rangle\otimes|\varphi\rangle$ and
$|\psi'\rangle\otimes|\varphi'\rangle$ of the product space is given
by:
\begin{equation}
(\langle\psi|\otimes\langle\varphi|)(|\psi'\rangle\otimes|\varphi'\rangle)=
\langle\psi|\psi'\rangle\langle\varphi|\varphi'\rangle.
\end{equation}
\noindent It is worth noting that when $\alpha$ and $\beta$ are
different, $|\alpha\rangle\otimes|\beta\rangle$ will be not the same vector as
$|\beta\rangle\otimes|\alpha\rangle$. Thus, in general, if $|\psi'\rangle$ and
$|\varphi'\rangle$ are linear combinations of basis vectors, it results
that:
\begin{equation}
(\langle\psi|\otimes\langle\varphi|)(|\psi'\rangle\otimes|\varphi'\rangle)\neq(\langle\psi|
\otimes\langle\varphi|)(|\varphi'\rangle\otimes|\psi'\rangle)
\end{equation}
\noindent and this has no sense when dealing with indistinguishable
particles. To solve this difficult, the symmetrization postulate is
assumed.

Our aim is to develop a procedure that takes into account
indistinguishability from the start, so we recall in which steps
artificial labeling has occurred. First of all, one assigns vectors
states to each particle in their corresponding Hilbert spaces and
``names'' in some way these spaces to perform the tensor product.
Informally, one says that one is making the product of the Hilbert
space of particle ``1" and the Hilbert space of particle ``2", and
does the same thing for the resulting states. As one arrives at a
situation in which particles seem to be distinguishable, one applies
a symmetrization postulate. Then, when defining the scalar product,
the differentiation of state spaces is maintained when taking
brackets, the bra of particle ``1" with the ket of particle ``1" and
the same for particle ``2". Thus, there are two steps to be avoided:
the use of the tensor product and this differentiation in the scalar
product.

To introduce the formal construction which will be developed in the
next sections, consider the possibility of a definition of a scalar
product resembling the following: Let $\{
|\alpha\rangle\otimes|\beta\rangle\}$ and
$\{|\alpha'\rangle\otimes|\beta'\rangle\}$ be two basis vectors of
the state space of the two particle system, then we could define
their scalar product as
\begin{equation}
(\langle\alpha|\otimes\langle\beta|)\circ(|\alpha'\rangle\otimes|\beta'\rangle)=
\delta_{\alpha\alpha'}\delta_{\beta\beta'}+\delta_{\alpha\beta'}\delta_{\beta\alpha'}.
\end{equation}
\noindent For any two vectors $|\psi\rangle\otimes|\varphi\rangle$
and $|\psi'\rangle\otimes|\varphi'\rangle$ that are linear
combinations of the basis ones, one would obtain:
\begin{equation}
(\langle\psi|\otimes\langle\varphi|)\circ(|\psi'\rangle\otimes|\varphi'\rangle)=
\langle\psi|\psi'\rangle\langle\varphi|\varphi'\rangle+\langle\psi|\varphi'\rangle\langle\varphi|\psi'\rangle.
\end{equation}
\noindent It is easy to verify that this product satisfies:
\begin{equation}
(\langle\psi|\otimes\langle\varphi|)\circ(|\psi\rangle\otimes|\varphi\rangle)=
|\psi|^{2}|\varphi|^{2}+|\langle\psi|\varphi\rangle|^{2}\geq
0,
\end{equation}
and also:
\begin{equation}
(\langle\psi'|\otimes\langle\varphi'|)\circ(|\psi\rangle\otimes|\varphi\rangle)=
((\langle\psi|\otimes\langle\varphi|)\circ(|\psi'\rangle\otimes|\varphi'\rangle))^*.
\end{equation}

\noindent where * stands for complex conjugation. Another
possibility to be considered is
\begin{equation}
(\langle\psi|\otimes\langle\varphi|)\bullet(|\psi'\rangle\otimes|\varphi'\rangle)=
\langle\psi|\psi'\rangle\langle\varphi|\varphi'\rangle-\langle\psi|\varphi'\rangle\langle\varphi|\psi'\rangle.
\end{equation}
This ``product'' clearly depends on the order of the terms, and it
is defined up to a minus sign. But recall that in quantum mechanics
we are interested in squared probability amplitudes and its square
does not depend on the order. Furthermore, the product $\bullet$ has
the following interesting property:
\begin{equation}
(\langle\psi|\otimes\langle\psi|)\bullet(|\psi\rangle\otimes|\psi\rangle)=
\langle\psi|\psi\rangle\langle\psi|\psi\rangle-\langle\psi|\psi\rangle\langle\psi|\psi\rangle=0,
\end{equation}
and this will turn of great importance, because if we interpret
$|\psi\rangle\otimes|\psi\rangle$ as an state with two fermions in
the same state, the state is a vector of null norm, and thus, null
probability, and also its scalar product with any other vector is
zero:
\begin{equation}
(\langle\varphi|\otimes\langle\phi|)\bullet(|\psi\rangle\otimes|\psi\rangle)=
\langle\phi|\psi\rangle\langle\varphi|\psi\rangle-\langle\phi|\psi\rangle\langle\varphi|\psi\rangle=0
\end{equation}
Moreover, using Cauchy-Schwartz inequality we have that
\begin{equation}
(\langle\psi|\otimes\langle\varphi|)\bullet(|\psi\rangle\otimes|\varphi\rangle)=
|\psi|^{2}|\varphi|^{2}-|\langle\psi|\varphi\rangle|^{2}\geq0,
\end{equation}
or
\begin{equation}
(\langle\psi|\otimes\langle\varphi|)\bullet(|\psi\rangle\otimes|\varphi\rangle)=
-|\psi|^{2}|\varphi|^{2}+|\langle\psi|\varphi\rangle|^{2}\leq0.
\end{equation}
These two possibilities come from the ambiguity in the sign when we
define ``$\bullet$". This ambiguity will be solved later.

\subsection{Construction of the \Q-space}\label{s:Construction}
In the following we apply the guiding ideas discussed above to
define a product in a \Q-\emph{space} constructed using the
quasi-set theory \Q.

\subsubsection{Quasi-functions}
Let us consider a \Q-set $\epsilon=_E\{\epsilon_{i}\}_{i \in I }$,
where $I$ is an arbitrary (denumerable or not) collection of
indexes, such that $Z(\epsilon)$. From now own, by a ``set" we mean
a \Q-set, and ``=" stands for ``$=_E$", except if explicitly
mentioned. We also recall that all the usual mathematical concepts
mentioned below can be obtained in the ``classical" part of \Q. We
take the elements $\epsilon_{i}$ to represent the eigenvalues of a
physical magnitude of interest. To fix the ideas, they may be the energy
eigenvalues of the Hamiltonian $H$ of the system,
$H|\varphi_{i}\rangle=\epsilon_{i}|\varphi_{i}\rangle$, with
$|\varphi_{i}\rangle$ being the corresponding eigenstates. The
construction we present is, of course, independent of this
particular choice. Consider then the quasi-functions $f$,
$f:\epsilon \longrightarrow \mathcal{F}_{p}$, where
$\mathcal{F}_{p}$ is the quasi-set formed of all finite and pure
quasi-sets. $f$ is the quasi-set formed of ordered pairs $\langle
\epsilon_{i};x\rangle$ with $\epsilon_{i}\in\epsilon$ and
$x\in\mathcal{F}_{p}$. Let us choice these quasi-functions in such a
way that whenever $\langle \epsilon_{i_{k}};x\rangle$ and $\langle
\epsilon_{i_{k'}};y\rangle$ belong to $f$ and $k\neq  k'$, then
$x\cap y=\emptyset$. Let us further assume  that the sum of the
quasi-cardinals of the quasi-sets which appear in the image of each
of these quasi-functions is finite, and then, $qc(x)=0$ for every
$x$ in the image of $f$, except for a finite number of elements of
$\epsilon$. Let us call $\mathcal{F}$ the quasi-set formed of these
quasi-functions. If $\langle x;\epsilon_{i}\rangle$ is a pair of
$f\in\mathcal{F}$, we will interpret that the energy level
$\epsilon_{i}$ has occupation number $qc(x)$. These quasi-functions
will be represented by symbols such as
$f_{\epsilon_{i_{1}}\epsilon_{i_{2}}\ldots\epsilon_{i_{m}}}$ (or by
the same symbol with permuted indexes). This indicates that the
levels $\epsilon_{i_{1}}\epsilon_{i_{2}}\ldots\epsilon_{i_{m}}$ are
occupied. It will be taken as convention that if the symbol
$\epsilon_{i_{k}}$ appears $j$-times, then the level
$\epsilon_{i_{k}}$ has occupation number $j$. For example, the
symbol
$f_{\epsilon_{1}\epsilon_{1}\epsilon_{1}\epsilon_{2}\epsilon_{3}}$
means that the level $\epsilon_{1}$ has occupation number $3$ while
the levels $\epsilon_{2}$ and $\epsilon_{3}$ have occupation numbers
$1$. The levels that do not appear have occupation number zero.

These quasi-functions will be used to construct quantum states. It
is worth to say that, because of the utilization of pure quasi-sets
with indistinguishable elements, there is no reference to particle
indexation. The only reference is to the occupation numbers, because
permutations make no sense here, as it should be. Let us consider,
for example, the quasi-function
$f_{\epsilon_{1}\epsilon_{1}\epsilon_{1}\epsilon_{2}\epsilon_{3}}$.
As we have said above, we interpret this as a state in which the
level $1$ has occupation number three, the levels $2$ and $3$ only
one, and the others zero. Thus, a permutation of particles makes no
change because the quasi-function
$f_{\epsilon_{1}\epsilon_{1}\epsilon_{1}\epsilon_{2}\epsilon_{3}}$
is a collection of ordered pairs. These pairs are
$\langle\epsilon_{1};x\rangle$, $\langle\epsilon_{2};y\rangle$,
$\langle\epsilon_{3};z\rangle$ and
$\langle\epsilon_{n};\emptyset\rangle$ (for $n>3$), where $x$, $y$
and $z$ are pure and disjoint quasi-sets which satisfy $qc(x)=3$ and
$qc(y) = 1 = qc(z)$. Thus, permutation of two particles is formally
represented by the procedure that takes an element of, say, $x$ and
interchanges it with an element of $y$ (or $z$). But it is a theorem
of \Q\ that permutation of $m$-atoms gives place to
indistinguishable quasi-sets (unobservability of permutations). By
definition, we have $\langle\epsilon_{1}; x\rangle =
[[\epsilon_{1}];[\epsilon_{1};x]]$. Also by definition,
$[\epsilon_{1};x]$ is the collection of all the indistinguishable
from either $\epsilon_{1}$ or $x$ (taken from some previously given
qset). For this reason, if we replace $x$ by $x'$, with $x\equiv x'$
we will obtain $[\epsilon_{1};x]=[\epsilon_{1};x']$. Thus, we obtain
$\langle\epsilon_{1};x\rangle= \langle\epsilon_{1};x'\rangle$ and
the ordered pairs of the `permuted' quasi-function will be the same
and, consequently, the new quasi-function is again
$f_{\epsilon_{1}\epsilon_{1}\epsilon_{1}\epsilon_{2}\epsilon_{3}}$.
We thus see that the permutation of indistinguishable elements does
not produce changes  in the quasi-functions and, then, in any vector
space constructed using them, the permutation operation will be
reduced to identity.

It is important to point out that the order of the indexes in a
quasi-function
$f_{\epsilon_{i_{1}}\epsilon_{i_{2}}\ldots\epsilon_{i_{n}}}$ has no
meaning at all because up to now, there is no need to define any
particular order in $\epsilon$, the domain of the quasi-functions of
$\mathcal{F}$. Nevertheless, we may define an order in the following
way. For each quasi-function $f\in\mathcal{F}$, let
$\{\epsilon_{i_{1}}\epsilon_{i_{2}}\ldots\epsilon_{i_{m}}\}$ be the
quasi-set formed by the elements of $\epsilon$ such that
$\langle\epsilon_{i_{k}},X\rangle\in f$ and $qc(X)\neq  0$ ($k=
1\ldots m$). We call $supp(f)$ this quasi-set (the \textit{support}
of $f$). Then consider the pair $\langle o,f\rangle$, where $o$ is a
bijective quasi-function:

$$o:\{\epsilon_{i_{1}}\epsilon_{i_{2}}\ldots\epsilon_{i_{m}}\}\longrightarrow
\{1,2,\ldots,m\}.$$

\noindent Each of these quasi-functions $o$ define an order on
$supp(f)$. For each $f\in\mathcal{F}$, if $qc(supp(f))= m$, then,
there are $m!$ orderings. Then, let $\mathcal{O}\mathcal{F}$ be the
quasi-set formed by all the pairs $\langle o,f\rangle$, where
$f\in\mathcal{F}$ and $o$ is a a particular ordering in $supp(f)$.
Thus, $\mathcal{O}\mathcal{F}$ is the quasi-set formed by all the
quasi-functions of $\mathcal{F}$ with ordered support. For this
reason, if we now say that
$f_{\epsilon_{i_{1}}\epsilon_{i_{2}}\ldots\epsilon_{i_{n}}}\in
\mathcal{O}\mathcal{F}$, we will be speaking of a quasifunction
$f\in\mathcal{F}$ and of an special ordering of
$\{{\epsilon_{i_{1}}\epsilon_{i_{2}}\ldots\epsilon_{i_{n}}}\}$. For
the sake of simplicity, we will use the same notation as before. But
now the order of the indexes {\it is meaningful}. It is also
important to remark, that the order on the indexes must not be
understood as a labeling of particles, for it easy to check that, as
above, the permutation of particles does not give place to a new
element of $\mathcal{O}\mathcal{F}$. This is so because a
permutation of particles operating on a pair $\langle
o,f\rangle\in\mathcal{O}\mathcal{F}$ will not change $f$, and so,
will not alter the ordering. We will use the elements of
$\mathcal{O}\mathcal{F}$ later, when we deal with fermions.

\subsubsection{Vector space structure}
A linear space structure is required to adequately represent quantum
states. To equip $\mathcal{F}$ and $\mathcal{OF}$ with such a
structure, we need to define two operations ``$\star$" and ``$+$", a
product by scalars  and an addition of their elements, respectively.
We will construct a vector space starting from the quasi-functions
of the quasi-sets $\mathcal{F}$ (or equivalently
$\mathcal{O}\mathcal{F}$) defined above. Call $C$ the collection of
quasi-functions which assign to every $f\in \mathcal{F}$ (or $f\in
\mathcal{O}\mathcal{F}$) a complex number. That is, a quasi-function
$c\in C$ is a collection of ordered pairs $\langle
f;\lambda\rangle$, where $f\in \mathcal{F}$ (or $f\in
\mathcal{O}\mathcal{F}$) and $\lambda$ a complex number. Let $C_{0}$
be the subset of $C$ such that, if $c\in C_0$, then $c(f)=0$ for
almost every $f\in \mathcal{O}\mathcal{F}$ (i.e., $c(f)=0$ for every
$f\in \mathcal{O}\mathcal{F}$ except for a finite number of
quasi-functions). We can define in $C_{0}$ a sum and a product by
scalars in the same way as it is usually done with functions as
follows.
\begin{definition}
Let $\alpha$, $\beta$ and $\gamma$ $\in \mathcal{C}$, and $c$,
$c_{1}$ and $c_{2}$ be  quasi-functions of $C_{0}$, then
$$(\gamma\ast c)(f) \igual \gamma(c(f))$$
$$(c_{1}+c_{2})(f) \igual  c_{1}(f) + c_{2}(f)$$
\end{definition}
\noindent The quasi-function $c_{0}\in C_{0}$ such that $c_{0}(f)=
0$, for any $f\in F$, acts as the null element of the sum, for
\begin{equation}
(c_{0}+c)(f)= c_{0}(f)+c(f)= 0+c(f)= c(f), \forall f.
\end{equation}
With the sum and the multiplication by scalars defined above we have
that $(C_{0},+,\ast)$ is a complex vector space. Each one of the
quasi-functions of $C_{0}$ should be interpreted in the following
way. If $c\in C_{0}$ (and $c\neq c_{0}$), let $f_{1}$, $f_{2}$,
$f_{3}$,$\ldots$, $f_{n}$ be the only functions of $C_{0}$ which
satisfy $c(f_{i})\neq  0$ ($i= 1,\ldots,n$). These quasi-functions
exist because, as we have said above, the quasi-functions of $C_{0}$
are zero except for a finite number of quasi-functions of
$\mathcal{F}$. If $\lambda_{i}$ are complex numbers which satisfy
that $c(f_{i})= \lambda_{i}$ ($i= 1,\ldots,n$), we will make the
association
\begin{equation}
c\approx(\lambda_{1}f_{1}+\lambda_{2}f_{2}+\cdots+\lambda_{n}f_{n}).
\end{equation}
The symbol $\approx$ must be understood in the sense that we use
this notation to represent the quasi-function $c$. The idea is that
the quasi-function $c$ represents the pure state which is a linear
combination of the states represented by the quasi-functions $f_{i}$
according to the interpretation given above. As a particular case of
this notation, we have that if $c_{j}\in C_{0}$ are the
quasi-functions such that  $c_{j}(f_{i})= \delta_{ij}$
($\delta_{ij}$ is the Kronecker symbol), then $c_{j}\approx f_{j}$
and in a similar way $\lambda\ast c_{j}\approx \lambda f_{j}$. In
this space, the vectors $c_{j}$ are the  ``natural" basis vectors,
while the others are linear combinations of them.

\subsubsection{Scalar products}
With the aid of a vector space structure, we can express quantum
superpositions. In order to calculate probabilities and mean values,
we need to introduce the notion of scalar product. In the following,
we will introduce two different products for bosons and fermions
separately, following the ideas of Section \ref{motivation}. Let us
do it first for bosons.

\begin{definition}
Let $\delta_{ij}$ be the Kronecker symbol and
$f_{\epsilon_{i_{1}}\epsilon_{i_{2}}\ldots\epsilon_{i_{n}}}$ and
$f_{\epsilon_{i'_{1}}\epsilon_{i'_{2}}\ldots\epsilon_{i'_{m}}}$ two
basis vectors, then
\begin{equation}\label{e:PS}
f_{\epsilon_{i_{1}}\epsilon_{i_{2}}\ldots\epsilon_{i_{n}}}\circ
f_{\epsilon_{i'_{1}}\epsilon_{i'_{2}}\ldots\epsilon_{i'_{m}}} \igual
\delta_{nm}\sum_{p}\delta_{i_{1}pi'_{1}}\delta_{i_{2}pi'_{2}}\ldots\delta_{i_{n}pi'_{n}}
\end{equation}
The sum is extended over all the permutations of the index set
$i'=(i'_{1},i'_{2},\ldots,i'_{n})$ and for each permutation $p$,
$pi'=(pi'_{1},pi'_{2},\ldots,pi'_{n})$.
\end{definition}
This product can be easily extended over linear combinations:
\begin{equation}\label{e:EXT}
(\sum_{k}\alpha_{k}f_{k})\circ(\sum_{k}\alpha'_{k}f'_{k}) \igual
\sum_{kj}\alpha_{k}^{\ast}\alpha'_{j}(f_{k}\circ f'_{j})
\end{equation}

On the other hand, we can consider another ``$\bullet$" product as
follows, which will be adequate for fermions:
\begin{definition}
Let $\delta_{ij}$ be the Kronecker symbol and
$f_{\epsilon_{i_{1}}\epsilon_{i_{2}}\ldots\epsilon_{i_{n}}}$ and
$f_{\epsilon_{i'_{1}}\epsilon_{i'_{2}}\ldots\epsilon_{i'_{m}}}$ two
basis vectors, then
\begin{equation}\label{e:PA}
f_{\epsilon_{i_{1}}\epsilon_{i_{2}}\ldots\epsilon_{i_{n}}}\bullet
f_{\epsilon_{i'_{1}}\epsilon_{i'_{2}}\ldots\epsilon_{i'_{m}}} \igual
\delta_{nm}\sum_{p}\sigma_{p}\delta_{i_{1}pi'_{1}}\delta_{i_{2}pi'_{2}}\ldots\delta_{i_{n}pi'_{n}}
\end{equation}
where:
\begin{equation}
\sigma_{p} = \cases {1, & if p is even  \cr -1 , & if p is odd \cr}
\end{equation}
\end{definition}
The result of this product is an antisymmetric sum of the indexes
which appear in the quasi-functions. In order that the product is
well defined, the quasi-functions must belong to
$\mathcal{O}\mathcal{F}$. Once this product is defined over the
basis functions, we can extend it to linear combinations, in a
similar way as in (\ref{e:EXT}). If the occupation number of a
product is more or equal than two, then the vector has null norm. As
it is a vector of null norm, the product of this vector with any
other vector of the space would yield zero, and thus the probability
of observing a system in a state like it vanishes. This means that
we can add to any physical state an arbitrary linear combination of
null norm vectors for they do not contribute to the scalar product,
which is the meaningful quantity.

We have defined two products, ``$\circ$" and ``$\bullet$", that are
adequate for bosons and fermions, respectively. We will return to
this point in the following section.

We point out that to formulate quantum mechanics in such a way that
no reference to particle individuality is made, we need to avoid
labeling in state vectors as much as in operators representing
observable quantities. As it is known, in the Fock-space formalism,
the observables can be written in terms of creation and annihilation
operators, avoiding particle labeling. In the following section we
will introduce creation and annihilation operators in order to
express observable quantities, without making appeal to particle
labeling in the operators themselves.

\section{The construction of quantum mechanics using \Q-spaces}\label{s:fock}

In this section we will first briefly review for completeness the
formulation of quantum mechanics using the Fock-space formalism.
After that, we will rewrite that formulation using the language of
the \Q-space.

\subsection{Fock-space formalism}\label{s:review}

As is well known, the standard formulation of quantum mechanics and
the Fock-space formulation are deeply connected. Equivalence with
wave mechanics is studied (for example) in \cite{rob73}. Here we
briefly recall some basic notions of the standard formalism to fix
notation. We call $T_{1}$ the kinetic energy of a single particle
and $V_{1}$ the external potential acting on it. For $n$ particles
we have:
\begin{equation}
T_{n} \igual  \sum_{i=1}^{n}T_{1}(r_{i})
\end{equation}
and the same for the external potential. We represent by
\begin{equation}
V_{n}\igual \sum_{i>j=1}^{n}V_{2}(\mathbf{r}_{i},\mathbf{r}_{j})
\end{equation}
the pairwise interaction potential. Thus, the total hamiltonian
operator is given by
\begin{equation}\label{e:nhamiltonian}
H_{n}=\sum_{i=1}^{n}[T_{1}(\mathbf{r}_{i})+V_{1}(\mathbf{r}_{i})+\sum_{i>j=1}^{n}
V_{2}(\mathbf{r}_{i},\mathbf{r}_{j})]
\end{equation}
The $n$-particles wave function is written as
\begin{equation}
\Psi_{n}(\mathbf{r}_{1},\ldots,\mathbf{r}_{n})
\end{equation}

The standard Fock-space is built up from the one particle Hilbert
spaces. Let $\mathcal{H}$ be a Hilbert space and define:
\begin{eqnarray}
&{\mathcal{H}}^{0}&= {\mathcal{C}}
\nonumber\\
&{\mathcal{H}}^{1}&= {\mathcal{H}}
\nonumber\\
&{\mathcal{H}}^{2}&= {\mathcal{H}}\otimes{\mathcal{H}}
\nonumber\\
&\vdots&\nonumber\\
&{\mathcal{H}}^{n}&= {\mathcal{H}}\otimes \cdots \otimes
{\mathcal{H}}
\end{eqnarray}
If no symmetry condition is required for the states, the Fock-space
is constructed as the direct sum of $n$ particles Hilbert spaces:
\begin{equation}
{\mathcal{F}}= \bigoplus^{\infty}_{n=0} {\mathcal{H}}^{n}
\end{equation}
When dealing with bosons or fermions, the standard procedure to
obtain the physical space state is as follows. Given a vector
$v=v_{1}\otimes\cdots\otimes v_{n}\in {\mathcal{H}}^{n}$, define:
\begin{equation}
\sigma^{n}(v)=(\frac{1}{n!})\sum_{P}P(v_{1}\otimes\cdots\otimes
v_{n})
\end{equation}
and:
\begin{equation}
\tau^{n}(v)=(\frac{1}{n!})\sum_{P}s^{p}P(v_{1}\otimes\cdots\otimes
v_{n})
\end{equation}
where:
\begin{equation}
s^{p} = \cases {1, & if p is even  \cr -1 , & if p is odd \cr}
\end{equation}
It is important to realize that in this construction, particles are
labeled and then symmetry conditions are imposed by state
symmetrization. Then, calling
\begin{equation}
{\mathcal{H}}^{n}_{\sigma}= \{\sigma^{n}(v):v\in {\mathcal{H}}^{n}
\}
\end{equation}
and:
\begin{equation}
{\mathcal{H}}^{n}_{\tau}= \{\tau^{n}(v):v\in {\mathcal{H}}^{n} \}
\end{equation}
we have the Fock-space
\begin{equation}
{\mathcal{F}}_{\sigma}= \bigoplus _{n=0}^{\infty}
{\mathcal{H}}^{n}_{\sigma}
\end{equation}
for bosons and
\begin{equation}
{\mathcal{F}}_{\tau}= \bigoplus^{\infty}_{n=0}
{\mathcal{H}}^{n}_{\tau}
\end{equation}
for fermions. Once each Fock-space is constructed, the usual
procedure runs as follows. Let $\psi(\mathbf{r})$ and its hermitian
conjugate $\psi(\mathbf{r})^{\dagger}$ be operators acting on the
Fock-space and satisfying:
\begin{eqnarray}
&[\psi(\mathbf{r}),\psi(\mathbf{r}')]_{\mp} = 0&
\nonumber\\
&[\psi(\mathbf{r})^{\dagger},\psi(\mathbf{r}')^{\dagger}]_{\mp} = 0&
\nonumber\\
&[\psi(\mathbf{r}),\psi(\mathbf{r}')^{\dagger}]_{\mp}=
\delta_{\mathbf{r}-\mathbf{r}'}&
\end{eqnarray}
\noindent where $\delta({\mathbf{r}}-{\mathbf{r}'})$ is the Dirac
delta function. For any operators $A$ and $B$, the brackets are
defined by:
\begin{equation}
[A,B]_{\mp} \igual AB\mp BA
\end{equation}
Corresponding to the $n$ particle wave function
$\Psi_{n}(\mathbf{r}_{1},\ldots,\mathbf{r}_{n})$ of the standard
formulation,  the $n$ particle state in the Fock-space is defined:
\begin{eqnarray}
|\psi_{n}\rangle \igual (n!)^{-\frac{1}{2}}\int d^{3}r_{1}\cdots\int
d^{3}r_{n}\psi(\mathbf{r}_{1})^{\dagger}\cdots\psi(\mathbf{r}_{n})^{\dagger}|0\rangle
\Psi_{n}(\mathbf{r}_{1},\ldots,\mathbf{r}_{n})
\end{eqnarray}
which can be shown to be an eigenvector (with eigenvalue $n$) of the
particle number operator:
\begin{equation}
N \igual \int d^{3}r \psi(\mathbf{r})^{\dagger}\psi(\mathbf{r})
\end{equation}
The connection between the two representations is given by:
\begin{equation}
\Psi_{n}(\mathbf{r}_{1},\cdots,\mathbf{r}_{n})=(n!)^{-\frac{1}{2}}\langle
0|\psi(\mathbf{r}_{1})\cdots\psi(\mathbf{r}_{n})|\Psi_{n}\rangle
\end{equation}
In general, an arbitrary vector of the Fock-space:
\begin{equation}
|\Psi\rangle=\sum_{n=0}^{\infty}|\Psi_{n}\rangle
\end{equation}
will not be an eigenstate of the particle number operator.

Corresponding to the kinetic energy operator of standard wave
mechanics, an operator in the Fock-space is defined as:
\begin{equation}
T\igual \int d^{3}r \psi^{\dagger}(\mathbf{r})T_{1}(r)\psi(\mathbf{r})
\end{equation}
and it is easy to see that:
\begin{equation}
T^{\dagger}=T
\end{equation}
It can also be shown that:
\begin{equation}
T|\Psi_{n}\rangle=(n!)^{-\frac{1}{2}}\int d^{3}r_{1}\cdots\int
d^{3}r_{n}\Psi^{\dagger}(\mathbf{r}_{1})\cdots\Psi^{\dagger}(\mathbf{r}_{n})
|0\rangle
\sum_{i=1}^{n}T_{1}(\mathbf{r}_{i})\Psi_{n}(\mathbf{r}_{1},\cdots,\mathbf{r}_{n})
\end{equation}
Analogously, if there is a pairwise interaction potential
$V_{2}(\mathbf{r},\mathbf{r}')$, the operator:
\begin{equation}
V \igual \frac{1}{2}\int d^{3}r\int
d^{3}r'\psi^{\dagger}(\mathbf{r})\psi^{\dagger}(\mathbf{r}')V_{2}(\mathbf{r},\mathbf{r}')
\psi(\mathbf{r}')\psi(\mathbf{r})
\end{equation}
is defined on the Fock-space. Its action on $|\Psi_{n}\rangle$ is
given by:
\begin{equation}
V|\Psi_{n}\rangle=(n!)^{-\frac{1}{2}}\int d^{3}r_{1}\cdots\int
d^{3}r_{1}[V,\psi^{\dagger}(\mathbf{r}_{n})\cdots\psi^{\dagger}(\mathbf{r}_{1})]
|\mathbf{0}\rangle\Psi_{n}(\mathbf{r}_{1}\ldots\mathbf{r}_{n})
\end{equation}
and it follows that:
\begin{eqnarray}
V|\Psi_{n}\rangle&=&(n!)^{-\frac{1}{2}}\int d^{3}r_{1}\cdots\int
d^{3}r_{1}\psi^{\dagger}(\mathbf{r}_{n})\cdots\psi^{\dagger}(\mathbf{r}_{1})
|\mathbf{0}\rangle\nonumber\\
&\times&\sum_{i=1}^{n}\sum_{j=1}^{i-1}V_{2}(\mathbf{r}_{i},\mathbf{r}_{j})
\Psi_{n}(\mathbf{r}_{1},\ldots,\mathbf{r}_{n})
\end{eqnarray}
It can be shown that that the following equations holds:
\begin{equation}
T_{n}\Psi_{n}(\mathbf{r}_{1},\cdots,\mathbf{r}_{n})=
(n!)^{-\frac{1}{2}}\langle
0|\Psi(\mathbf{r}_{1})\cdots\Psi(\mathbf{r}_{n})T|\Psi_{n}\rangle
\end{equation}
\begin{equation}
V_{n}\Psi_{n}(\mathbf{r}_{1},\cdots,\mathbf{r}_{n})=
(n!)^{-\frac{1}{2}}\langle
0|\Psi(\mathbf{r}_{1})\cdots\Psi(\mathbf{r}_{n})V|\Psi_{n}\rangle
\end{equation}
where:
\begin{equation}
T_{n} \igual \sum_{i=1}^{n}T_{1}(\mathbf{r}_{i})
\end{equation}
\begin{equation}
V_{n}\igual \sum_{i>j=1}^{n}V_{2}(\mathbf{r}_{i},\mathbf{r}_{j})
\end{equation}

The equivalence with wave mechanics can now be established as
follows. If $\Psi_{n}(\mathbf{r}_{1},\cdots,\mathbf{r}_{n})$
satisfies the $n$ particle Schr\"{o}dinger wave equation with
Hamiltonian (\ref{e:nhamiltonian}), it follows that in the
Fock-space formulation $|\Psi_{n}\rangle$ must satisfy the
Fock-space Schr\"{o}dinger equation:
\begin{equation}
[i\hbar(\frac{\partial}{\partial t})-H]|\Psi_{n}\rangle=0
\end{equation}
with $H \igual T+V$: given by:
\begin{eqnarray}\label{e:hamiltonian}
H&=&\int d^{3}r
\psi^{\dagger}(\mathbf{r})[T_{1}+V_{1}(\mathbf{r})]\psi(\mathbf{r})
\nonumber\\
&+&\frac{1}{2}\int d^{3}r\int
d^{3}r'\psi^{\dagger}(\mathbf{r})\psi^{\dagger}(\mathbf{r}')V_{2}(\mathbf{r},\mathbf{r}')
\psi(\mathbf{r}')\psi(\mathbf{r})
\end{eqnarray}
It is important to remark that the $n$ particle Schr\"{o}dinger wave
equation is not completely equivalent to its analogue in the
Fock-space formalism. Only solutions of the Fock-space equation
which are eigenvectors of the particle number operator with particle
number $n$ can be solutions of the corresponding $n$ particle
Schr\"{o}dinger wave equation. On the other hand, not all the
solutions of the $n$ particle Schr\"{o}dinger wave equation can be
solutions of the Fock equation, only those which are adequately
symmetrized do. So, both conditions, defined particle number and
symmetrization, must hold in order that both formalisms are
equivalent.

\subsection{Creation and annihilation operators}\label{s:Creation}
The standard manner to handle with the equations in Fock-space is to
write physical magnitudes in terms of creation and annihilation
operators. To do so, one makes the following expansion:
\begin{equation}\label{e:decom}
\psi(\mathbf{r})=\sum_{k}a_{k}u_{k}(\mathbf{r})
\end{equation}
The coefficients of the expansion are the annihilation operators:

\begin{equation}
a_{k} \igual \int d^{3}ru_{k}^{\ast}(\mathbf{r})\psi(\mathbf{r})
\end{equation}
A similar expansion stands for the creation operator:
$a_{k}^{\dagger}$. In quantum field theory, it is commonly assumed
that the action of the operator $a_{k}^{\dagger}$ describes the
``creation of a particle" with wave function $u_{k}(\mathbf{r})$. In
a similar way, is interpreted that the action of $a_{k}$ describes
the ``annihilation of a particle". It can be shown, that these
operators satisfy the commutation relations:

\begin{eqnarray}
&[a_{k},a_{l}]_{\mp} = 0&
\nonumber\\
&[a_{k}^{\dagger},a_{l}^{\dagger}]_{\mp} = 0&
\nonumber\\
&[a_{k},a_{l}^{\dagger}]_{\mp}= \delta_{kl}&
\end{eqnarray}

\begin{equation}
N_{k} \igual a_{k}^{\dagger}a_{k}
\end{equation}

\noindent We can cast these last equations in a more familiar form,
using the ``$[,]$" symbol for bosonic commutation relations and the
``$\{,\}$" symbol for fermionic (anti)commutation relations. Then
for bosons we have:

\begin{equation}\label{e:BOS1}
[a_\alpha;a^{\dag}_\beta]=a_\alpha a^{\dag}_\beta-a^{\dag}_\beta
a_\alpha=\delta_{\alpha\beta}I
\end{equation}
\begin{equation}\label{e:BOS1}
[a^{\dag}_\alpha;a^{\dag}_\beta]=0
\end{equation}
\begin{equation}\label{e:BOS1}
[a_\alpha;a_\beta]=0
\end{equation}
and for fermions, (with $C^{\dag}_\alpha$ and $C_\alpha$ playing the
role of fermionic creation and annihilation operators respectively)
we have:
\begin{equation}\label{e:FER2}
\{C_\alpha;C^{\dag}_\beta\}=C_\alpha C^{\dag}_\beta+C^{\dag}_\beta
C_\alpha=\delta_{\alpha\beta}I
\end{equation}
\begin{equation}\label{e:FER3}
\{C^{\dag}_\alpha;C^{\dag}_\beta\}=0
\end{equation}
\begin{equation}\label{e:FER4}
\{C_\alpha;C_\beta\}=0
\end{equation}
Substitution of (\ref{e:decom}) in (\ref{e:hamiltonian}) yields:

\begin{equation}
H=\sum_{kl}a_{k}^{\dagger}T_{kl}a_{l}+\frac{1}{2}\sum_{klpq}a_{k}^{\dagger}a_{l}^{\dagger}V_{klpq}
a_{p}a_{q}
\end{equation}
where the matrix elements $T_{kl}$ and $V_{klpq}$ are given by:

\begin{eqnarray}
&T_{kl}=\int
d^{3}ru_{k}^{\ast}(\mathbf{r})[(-\frac{\hbar^{2}\nabla^{2}}{2m})+V_{1}(\mathbf{r})]u_{l}(\mathbf{r})&
\nonumber\\
&V_{klpq}=\int d^{3}r\int
d^{3}r'u_{k}^{\ast}(\mathbf{r})u_{l}^{\ast}(\mathbf{r}')V_{2}(\mathbf{r},\mathbf{r}')u_{p}(\mathbf{r}')
u_{q}(\mathbf{r})&
\end{eqnarray}

\noindent and similar expressions can be found for more general
obserables.
\subsection{Using the \Q-space}\label{s:Using}
We have constructed two spaces whose vectors make no reference to
particle indexation and, besides, particles are not labeled in any
step of the formal construction. This is possible because these
spaces are constructed using the non classical part of \Q, which may
refer to intrinsically indistinguishable entities. Vectors in these
spaces are only distinguished by the occupation number in each
(energy) level. With these tools and using the language of \Q, the
formalism of quantum mechanics may be completely rewritten giving a
straightforward answer to the problem of giving a formulation of
quantum mechanics in which intrinsical indistinguishability is taken
into account from the beginning, without artificially introducing
extra postulates.

Let us first show that the \Q-space is useful to provide a states
space analogous to  Fock-space. With this aim, we make the following
association in order to turn the notation similar to that of
standard quantum mechanics. For each quasi-function
$f_{\epsilon_{i_{1}}\epsilon_{i_{2}}\ldots \epsilon_{i_{n}}}$ of the
quasi-sets $\mathcal{F}$ or $\mathcal{O}\mathcal{F}$ constructed
above, we will write:

$$\alpha f_{\epsilon_{i_{1}}\epsilon_{i_{2}}\ldots\epsilon_{i_{n}}}\igual
\alpha|\epsilon_{i_{1}}\epsilon_{i_{2}}\ldots \epsilon_{i_{n}})$$

\noindent with the obvious corresponding generalization for linear
combinations.

Let us recall again that in $|\epsilon_{i_{1}}\epsilon_{i_{2}}\ldots
\epsilon_{i_{n}})\in \mathcal{F}$, the order of the indexes has no
meaning. But in $|\epsilon_{i_{1}}\epsilon_{i_{2}} \ldots
\epsilon_{i_{n}})\in \mathcal{O}\mathcal{F}$, the order makes sense.

As we have already pointed out that, to avoid particle labeling in
the expressions for observables, no reference to particle indexation
should appear in their corresponding operators. For that reason we
will only use creation and annihilation operators. To do so, we
construct creation and annihilation acting on \Q-spaces. In the rest
of this section, we will develop the idea of the symmetrized
products (\ref{e:PS}) y (\ref{e:PA}) discussed in
\ref{s:Construction}. We will first develop the construction for
bosons and later fermions. We will use creation and annihilation
operators and instead of postulating commutation relations, we will
deduce them from their definitions and the properties of the vectors
of the \Q-spaces (following an analogous procedure as that exposed,
for example, in \cite[Chap.\ 17]{bal00}).

\subsection{Bosonic states}\label{s:bosons}
For bosons, the procedure is similar to the procedure of the
standard approach for, as we have remarked earlier, a scalar product
naturally arises from the symmetric product (\ref{e:PS}). This
implies that, once normalized to unity, the vectors
$|\alpha\beta\gamma\ldots)$ constructed using \Q, are equivalent to
the symmetrized vectors $|\alpha\beta\gamma\ldots\rangle$ for
bosonic states. This is so, because permutations alter the vector in
none of the spaces.

Suppose then that vectors $|\alpha\beta\gamma\ldots)$ are normalized to
unity. If $\zeta$ represents an arbitrary collection of indexes, we
define:
\begin{equation}
a^{\dag}_\alpha|\zeta)\propto|\alpha\zeta)
\end{equation}
in such a way that the proportionality constant satisfies
\begin{equation}\label{NB}
a^{\dag}_\alpha a_\alpha|\zeta)=n_{\alpha}|\zeta)
\end{equation}
Then, it follows that:
\begin{equation}
((\zeta|a^{\dag}_\alpha)( a_\alpha|\zeta))=n_{\alpha}
\end{equation}
obtaining the usual
\begin{definition}
\begin{equation}
a_\alpha|\ldots n_{\alpha}\ldots)=\sqrt{n_{\alpha}} \; |\ldots n_{\alpha}-1\ldots)
\end{equation}
\end{definition}
On the other hand,
\begin{equation}
a_\alpha a^{\dag}_\alpha
|\ldots n_{\alpha}\ldots)=K\sqrt{n_{\alpha}+1} \; |\ldots n_{\alpha}\ldots)
\end{equation}
where $K$ is a proportionality constant. If we apply
$a^{\dag}_\alpha$ once again:
\begin{equation}
a^{\dag}_\alpha a_\alpha a^{\dag}_\alpha
|\ldots n_{\alpha}\ldots)=K^{2}\sqrt{n_{\alpha}+1} \; |\ldots n_{\alpha}+1\ldots)
\end{equation}
and using (\ref{NB}):
\begin{equation}
(a^{\dag}_\alpha a_\alpha)a^{\dag}_\alpha
|\ldots n_{\alpha}\ldots)=\sqrt{n_{\alpha}+1}K|\ldots n_{\alpha}+1\ldots)
\end{equation}
so $K=\sqrt{n_{\alpha}+1}$. Then,
\begin{definition}
\begin{equation}
a^{\dag}_\alpha|\ldots n_{\alpha}\ldots)=\sqrt{n_{\alpha}+1} \; |\ldots n_{\alpha}+1\ldots)
\end{equation}
\end{definition}
Once this is established, let us obtain the commutation relations.
By a straightforward computation, we see that:
\begin{equation}
(a_\alpha a^{\dag}_\beta-a^{\dag}_\beta
a_\alpha)|\psi)=\delta_{\alpha\beta}|\psi)
\end{equation}
which is the same as:
\begin{equation}
[a_\alpha;a^{\dag}_\beta]=\delta_{\alpha\beta}I
\end{equation}
In an analogous way we can show that:
\begin{equation}
[a_\alpha;a_\beta]=[a^{\dag}_\alpha;a^{\dag}_\beta]=0
\end{equation}
This shows that the (bosonic) commutation relations that are
obtained in \Q-space are the same ones as in the standard
Fock-space.

\subsection{Fermionic states}\label{s:fermions}
For the fermionic case, we will use $C_{0}$ equipped  with the
antisymmetric product given by equation (\ref{e:PA}). We define the
creator operator $C^{\dag}_\alpha$ as follows:
\begin{definition}\label{e:DEF} Let $\zeta$ represent a collection of
indexes with non null occupation number, then
\begin{equation}
C^{\dag}_\alpha|\zeta)=|\alpha\zeta)
\end{equation}
\end{definition}
Note that if $\alpha$ was already in the collection $\zeta$, then
$|\alpha\zeta)$ is a vector with null norm. To have null norm
implies that $(\psi|\alpha\zeta)=0$ for all $|\psi)$. Then, if a
given vector has null norm, its scalar product with any other vector
in the space is zero. It follows that in the case that a system were
eventually in a state of null norm, the probability of observing it
would be zero. In the same way, if a linear combination of null norm
vectors were added to the vector representing the state of a system,
this addition would not give place to observable results. It follows
then that null norm vectors do not represent real physical states,
and the same holds for linear combinations of them. Moreover, adding
a vector of null norm to any other one does not produce observable
affects, because the terms of null norm do not contribute to the mean
values or to the probabilities. In order to express this situation, we
define the following relation:
\begin{definition}\label{e:chirimbolo} Two vectors $|\varphi)$ and $|\psi)$
are similar (and we will write $ |\varphi)\cong|\psi)$)) if the
difference between them is a linear combination of null norm
vectors.
\end{definition}
Let us now compute the effect of applying $C_{\alpha}$ to the
vectors of $\mathcal{O}\mathcal{F}$. Using Definition (\ref{e:DEF})
we find that:
\begin{equation}
(\zeta|C_\alpha=(\alpha\zeta|
\end{equation}
Then, for any vector $|\psi)$:
\begin{equation}\label{e:miga}
(\zeta|C_\alpha|\psi)=(\alpha\zeta|\psi)=0
\end{equation}
for $\alpha\in\zeta$ or $(\psi|\alpha\zeta)=0$. Then, if we choose
$|\psi)=|0)$ it follows that:
\begin{equation}
(\zeta|C_\alpha|0)=(\alpha\zeta|0)=0
\end{equation}
and thus we obtain that $C_\alpha|0)$ is orthogonal to any vector
which contains $\alpha$ and to any vector which does not contain
$\alpha$. Then, it is orthogonal to any vector, and for that reason,
it has to be  a linear combination of null norm vectors. Then, we do
not loose anything if we establish $C_\alpha|0)=\vec{0}$. In an
analogous way we can assert that:
\begin{equation}
C_\alpha|(\sim\alpha)\cdots)=\vec{0}
\end{equation}
where $(\sim\alpha)$ means that $\alpha$ has occupation number zero,
and the dots mean that the other levels have arbitrary occupation
numbers. Using (\ref{e:chirimbolo}) we can also write:
\begin{equation}
C_\alpha|0)\cong\vec{0}
\end{equation}
and
\begin{equation}
C_\alpha|(\sim\alpha)\ldots)\cong\vec{0}
\end{equation}
In what follows we will use $\cong$ when it be necessary, but the
same results are obtained if we replace $\cong$ by the extensional
equality. Making $|\psi)=|\alpha)$ in (\ref{e:miga}) it follows
that:
\begin{equation}
(\zeta|C_\alpha|\alpha)=(\alpha\zeta|\alpha)=0
\end{equation}
in any case except for $|\zeta)=|0)$. In that case,
$(0|C_\alpha|\alpha)=1$. Then, it follows that
$C_\alpha|\alpha)\cong|0)$. In an analogous way we obtain:
\begin{equation}
C_\alpha|\alpha\zeta)\cong|(\sim\alpha)\zeta)
\end{equation}
if $\alpha$ does not belongs to $\zeta$. But in the case that
$\alpha$ belongs to $\zeta$, (i.e., the occupation number is greater
than $1$) we have that $|\alpha\zeta)$ has null norm, and so:
\begin{equation}
(\alpha\zeta|C^{\dag}_\alpha|\psi)=(\alpha\zeta|\alpha\psi)=0,\forall|\psi)
\end{equation}
From this equation it follows that:
\begin{equation}
(\psi|C_\alpha|\alpha\zeta)=0,\forall|\psi)
\end{equation}
and so, $C_\alpha|\alpha\zeta)$ has null norm too.

Now, let us find the anti-commutation relations obeyed by the
fermionic creation and annihilation operators.
Let us first calculate the commutation relation between $C_\alpha$
and $C^{\dag}_{\beta}$. To do so, let us first study the
relationship between $|\alpha\beta)\in\mathcal{O}\mathcal{F}$ and
$|\beta\alpha)\in\mathcal{O}\mathcal{F}$. With this aim, consider
the vector $ |\alpha\beta)+|\beta\alpha)$ and perform the product of
this sum with another arbitrary vector. It suffices to study what
happens with basis vectors. The product yields trivially zero for
any vector different from $|\alpha\beta)$ or $|\beta\alpha)$. Making
the product with $|\alpha\beta)$ we obtain:
\begin{eqnarray}
&(\alpha\beta|[|\alpha\beta)+|\beta\alpha)]=
(\alpha\beta||\alpha\beta)+(\alpha\beta||\beta\alpha)=&\nonumber\\
&\delta_{\alpha\alpha}\delta_{\beta\beta}-
\delta_{\alpha\beta}\delta_{\beta\alpha}+\delta_{\alpha\beta}\delta_{\alpha\alpha}
-\delta_{\alpha\alpha}\delta_{\beta\beta}=1-0+0-1=0&
\end{eqnarray}
The same conclusion holds if we multiply it by $|\beta\alpha)$. Then,
it follows that $|\alpha\beta)+|\beta\alpha)$ is a linear
combination of null norm vectors (which we will denote by $|nnlc)$)
and, thus:
\begin{equation}
|\alpha\beta)=-|\beta\alpha)+|nnlc)
\end{equation}
We do not care about which is the particular null norm linear
combination, because it has no observable effects. Now, we can
calculate
\begin{equation}
C^{\dagger}_\alpha
C^{\dagger}_\beta|\psi)=|\alpha\beta|\psi)=_-|\beta\alpha\psi)+|nnlc)=
-C^{\dagger}_\beta C^{\dagger}_\alpha|\psi)+|nnlc)
\end{equation}
and thus
\begin{equation}
\{C^{\dag}_\alpha;C^{\dag}_\beta\}|\psi)=|nnlc)
\end{equation}
Then, we do not loose generality if we set
\begin{equation}
\{C^{\dag}_\alpha;C^{\dag}_\beta\}|\psi)= 0
\end{equation}
In an analogous way, we conclude that
\begin{equation}
\{C_\alpha;C_\beta\}|\psi)= 0
\end{equation}
Now let us calculate the commutation relation between $C_\alpha$ and
$C^{\dag}_\beta$. Suppose first that $\alpha\neq\beta$. If
$\alpha\notin\psi$ or $\beta\in\psi$ then it is clear that
\begin{equation}
\{C_\alpha;C^{\dag}_\beta\}|\psi)\approx \overrightarrow{0}
\end{equation}
If $\alpha\in\psi$ and $\beta\notin\psi$, suppose (without loss of
generality), that $\alpha$ is the first symbol in the list of
$\psi$. Then,
\begin{eqnarray}
&\{C_\alpha;C^{\dag}_\beta\}|\psi)=
C_\alpha|\beta\psi)+C^{\dag}_\beta|\psi(\sim\alpha))\cong&\nonumber\\
&\cong-|\beta\psi(\sim\alpha))+|\beta\psi(\sim\alpha))=
\overrightarrow{0}&
\end{eqnarray}
If $\alpha=\beta$, and $\alpha\in\psi$, then
\begin{eqnarray}
&\{C_\alpha;C^{\dag}_\alpha\}|\psi)=
C_\alpha|\alpha\psi)+C^{\dag}_\alpha|\psi(\sim\alpha))\cong&\nonumber\\
&\cong\overrightarrow{0}+|\psi)= |\psi)&
\end{eqnarray}
If $\alpha=\beta$, and $\alpha\notin\psi$, then
\begin{eqnarray}
&\{C_\alpha;C^{\dag}_\alpha\}|\psi)=
C_\alpha|\alpha\psi)+C^{\dag}_\alpha|\psi(\sim\alpha))\cong&\nonumber\\
&\cong|\psi)+\overrightarrow{0}= |\psi)&
\end{eqnarray}
So, in any case, we recover the relation
\begin{equation}
\{C_\alpha;C^{\dag}_\alpha\}|\psi)\cong\delta_{\alpha\beta}|\psi)
\end{equation}
and then, we can set
\begin{equation}
\{C_\alpha;C^{\dag}_\alpha\}=\delta_{\alpha\beta}.
\end{equation}
Thus, we have shown that the same commutation relations hold, as  the
standard formalism hold in \Q-space. This means that both
formulations are equivalent, for all the interesting information is
contained in the commutation relations. In the following section, we
discuss some features of this new formulation.

\section{Discussion}\label{s:discussion}
We have shown that it is possible to construct the quantum
mechanical formalism for indistinguishable particles without
labeling them in any step. To do so, we have built a vector space
with inner product, the \Q-space, using the non-classical part of
\Q, the generalization of ZFU, to deal with indistinguishable
elements. Vectors in \Q-space refer only to occupation numbers and
permutations operators act as the identity operator, reflecting in
the formalism the fact of unobservability of permutations, already
expressed in terms of the formalism of \Q.

We have also argued that it is useful to represent operators (which
are intended to represent observable quantities) as combinations of
creator and annihilation operators, in order to avoid particle
indexation in the expression of observable quantities. We have shown
that creation and annihilation operators which act on \Q-space can
be constructed. We have proved that they obey the usual commutation
and anticommutation relations for bosons and fermions respectively,
and this means that our construction is equivalent to that of the
Fock-space formulation of quantum mechanics. Thus, using the results
reviewed in section \ref{s:fock}, this implies that we can recover
the $n$-particles wave equation using \Q-space in the same way as in
the standard theory. Though both formulations are equivalent `for
all practical purposes', when subjected to careful analysis, the
conceptual difference turns very important. Our construction avoids
the LTPSF by constructing the state spaces using \Q, a theory which
can deal with truly indistinguishable entities, and so, it gives an
alternative (and radical) answer to the problems posed in
\cite{redtel92}, so as (we guess) answers Manin's problem posed in \cite{man76}.

This last point seems remarkable, for our construction incorporates
intrinsical indistinguishability from the beginning. Thus, our
approach fulfills not only Post's claim already mentioned, but also
both Manin's claim that we should find an adequate ``set theory" for
expressing collections of indistinguishable quanta, and Heisenberg's
idea that new tools (perhaps new logical tools) seem to be justified
in approaching ``new fields of experience", as we see in the
quotation at the beginning of our Introduction (which we invite the
reader to have a new look), so as also the following quotation from Manin:

\begin{quote}
``In accordance with Hilbert's prophecy, we are living in Cantor's
paradise. So we are bound to be tempted. $\ldots$

``We should consider the possibilities of developing a totally new
language to speak about infinity.\footnote{Set theory is also known
as the theory of the `infinite'.} Classical cri\-tics of Can\-tor
(Brou\-wer \textit{et al.}) argued that, say, the general choice
axiom is an illicit extrapolation of the finite case.

 ``I would like to point out that this is rather an
extrapolation of common-place physics, where we can distinguish
things, count them, put them in some order, etc. New quantum physics
has shown us models of entities with quite different behavior. Even
`sets' of photons in a looking-glass box, or of electrons in a
nickel piece are much less Cantorian than the `set' of grains of
sand. In general, a highly probabilistic `physical infinity' looks
considerably more complicated and interesting than a plain infinity
of `things'.''
\end{quote}

Manin is right. The foundation analysis of a living science is of course a difficult
problem. As far as we go to the details, science itself changes, and
in certain sense axiomatization becomes (as it was considered long
time ago) just a cosmetics to the scientific theories. But this
conclusion does not make justice to the advance of the modern
techniques of logic and mathematics. A careful look to the foundational
details may illuminate conceptual problems, open new windows and
show new mathematical and logical questions. Once more referring to
Yuri Manin, we think he is right in saying that ``The twentieth
century return to Middle Age scholastics taught us a lot about
formalisms. Probably it is time to look outside again. Meaning is
what really matters'' \cite{man76}. Quantum physics is a wonderful
land to look at. As our author also says,

\begin{quote}
``The development of the foundations of physics
in the twentieth century has taught us a serious lesson. Creating
and understanding these  foundations turned out to have very little
to do with the epistemological  abstractions which were of such
importance to the twentieth century critics of the foundations of
mathematics: finiteness, consistency, constructibility, and, in
general, the Cartesian notion of intuitive clarity. Instead,
completely unforeseen principles moved into the spotlight:
complementarity, and the nonclassical, probabilistic truth function.
The electron is infinite, capricious, and free, and does not at all
share our love for algorithms.'' \cite[pp.\ 82-83]{man77}
\end{quote}


\end{document}